\documentclass[12pt]{article}

\newcommand{\be}{\begin{eqnarray}}
\newcommand{\ee}{\end{eqnarray}}
\newcommand{\p}{\partial}
\newcommand{\ve}{\varepsilon}
\newcommand{\rhob}{{\bar \rho}}
\newcommand{\lra}{\longrightarrow}
\newcommand{\lt}{\tilde{\lambda}}
\newcommand{\e}{\epsilon}
\newcommand{\mut}{{\tilde \mu}}
\newcommand{\phit}{\tilde \phi}
\newcommand{\at}{\tilde a}

\def\gtrsim{\ \raisebox{-.6ex}{\rlap{$\sim$}} \raisebox{.4ex}{$>$}\ }

\usepackage{psfig}

\begin{document}

\vspace* {-35 mm}
\begin{flushright}  OSLO-TP 10-97 \\
November-1997
\end{flushright}

\vskip 20mm
\centerline{\Large\bf Bulk and edge properties}
\centerline{\Large\bf of the Chern-Simons Ginzburg-Landau theory}
\centerline{\Large\bf for the fractional quantum Hall effect}

\vskip 30mm
\centerline{\bf J.M. Leinaas and S. Viefers}               
\begin{center}
{\em Department of Physics \\
P.O. Box 1048 Blindern \\
N-0316 Oslo \\
Norway}
\end{center}
\vskip 15mm
\newcommand \sss {\mbox{ $<\overline{s}s>$} }
\def\fk{\mbox{ $f_K$} }
\centerline{\bf ABSTRACT}
\vskip 3mm
The Chern-Simons Ginzburg-Landau theory for the fractional Quantum Hall effect  
is studied in the presence of a confining potential. We review the bulk
properties of
the model and discuss how the  plateau formation emerges without any
impurity potential.
The effect is related to changes, by accumulation of charge, at the edge
when the
chemical potential is changed. Fluctuations about the ground state are
examined and
an expression is found for the velocity of the massless edge mode in terms
of the
confining potential. The effect of including spin is examined for the case
when the
system is fully polarized in the bulk. In general a spin texture may appear at
the edge,
and we examine this effect in the case of a small spin down component. The
low frequency
edge modes are examined and a third order equation is found for velocities which
indicates the presence of three different modes. The discussions are
illustrated by
numerical studies of the ground states, both for the one- and two-component
cases.

\vfil
\noindent

\eject
\section{Introduction}

Since the experimental discovery of the quantum Hall effect (QHE)
\cite{vonklitzing1,tsui1}, which occurs in a two-dimensional electron gas
subject to a
strong perpendicular magnetic field, several approaches to a theoretical
description of
this system have been developed. As a useful supplement to existing microscopic
descriptions \cite{laughlin1}, Zhang, Hansson and Kivelson \cite{zhang1} in 1988
proposed a Chern-Simons Ginzburg-Landau (CSGL) model for the fractional
quantum Hall
effect (FQHE). This model is based on the concept of ``statistical
transmutation'',
{\em i.e.} the fact that in two dimensions, fermions can be described as
(charged)
bosons carrying an odd integer number of (statistical) flux quanta. At the
Lagrangian
level this is done by adding Chern-Simons fields coupled to the bosons. In
this way,
electrons in an external magnetic field are described as bosons in a
combined external
and statistical magnetic field. At special values of the filling fraction the
statistical field cancels the external field (in the mean field sense) and
the system
is described as a gas of bosons feeling no net magnetic field. These bosons
``condense'' into a homogeneous ground state. In fact, this rather 
simple effective
theory reproduces several of the key features of the FQHE \cite{zhang1},
such as the
quantization of the Hall conductance and the existence of (anti)vortex
excitations. It
is even possible to re-construct Laughlin's electron wave functions from
this model by
considering quantum fluctuations about the (mean field) boson ground state
\cite{zhang2}.

An interesting aspect of the quantum Hall system is the existence of
gapless edge modes
\cite{wen1} which has important consequences for the transport properties
of the system
\cite{TRANSPORT}. Several authors have studied the CSGL model in the
presence of an
(infinitely) steep external confining potential and have shown the presence
of gapless
edge modes \cite{nagaosa1,orgad1,morinari1} in this model. The description
of these
modes in the CSGL model has also been related \cite{orgad2} to the chiral
Luttinger
liquid description of the edge excitations
\cite{haldane1,wen1}.

In this paper we study a number of aspects of the CSGL model, discussing
both its
ground state and excitations. We focus, in particular, the attention on
edge effects
and follow up the discussion of previous papers on this subject by
considering the
density profile and edge excitations for a smooth confining potential. We also
examine the effects on the edge of including the electron spin in the
description.

A short presentation of the model in its simplest form is given in section
\ref{modelsec}. It describes the quantum Hall system without spin at
Laughlin fillings
$\nu=1/(2m+1)$. In section \ref{bulksec}, we review the ground state
properties of the bulk and discuss how the existence of finite-energy
vortex excitations
leads to incompressibility of the ground state and the existence of 
quantum Hall
plateaux, two key features of the QH system. The edge of the system in the
presence of a smooth confining potential is then studied (section
\ref{edgesec}). In the
ground state (with uniform bulk density) one finds a one-parameter family
of solutions
parametrized by the total charge at the edge \cite{orgad1}. When this
charge exceeds
an upper critical value, the ground state with uniform bulk density becomes
unstable due
to formation of anti-vortices. Similarly, there is a lower critial value of
the total
charge beyond which the bulk tends to rearrange itself by formation of
vortices. A
numerical study of the ground state within these limits is presented. 
We further
perform a systematic analysis of edge excitations and show the presence of
a massless
mode with dispersion of the expected form  $\omega/q = <E>/B$ \cite{wen1},
and with an
explicit expression for the averaged electric field $<E>$.

In section \ref{twocompsec} we generalize the model to include spin by
introducing a
two-component description with a Zeeman interaction. The ground state is
completely
polarized in the bulk, however, a spin texture at the edge may be energetically
favourable if the Zeeman energy is not too large. We study the ground state
analytically and numerically; there is a two-parameter family of solutions
parametrized
by two conserved charges. The numerical analysis provides estimates of the
critical
Zeeman gap and it shows rotation of the spin vector along the edge for
smaller values
of the Zeeman energy. (The same effect has been demonstrated in a recent
study by
Karlhede, Kivelson, Lejnell and Sondhi \cite{karlhede1} who used
Hartree-Fock techniques
and an effective spin model.) We further examine the edge excitations of the
spin-polarized system and find an equation for the mode frequencies which
indicates the
presence of three massless modes at the edge.

Finally, in section \ref{discsec} we summarize and discuss our results.

\section{The model}              \label{modelsec}

The spinless system is described by the Lagrangian \cite{zhang1}
\be
{\cal L} = \phi^*(i\p_t - a_0 - eA_0 - V({\bf x}) + \mu) \phi
         - \kappa\left| \left( \nabla-i({\bf a}+e{\bf A}) \right) 
           \phi \right|^2
         - \lambda \left| \phi \right|^4
         + \frac{1}{4\theta} \ve^{\mu\nu\lambda}a_{\mu}\p_{\nu}a_{\lambda}.
                                        \label{model}
\ee
The Bose field $\phi$ is coupled both to an external field ${\bf A}=(0,Bx)$ with 
$B<0$,
and to
the statistical field ${\bf a}$.  $\mu$ is a chemical potential, $e=-|e|$ 
denotes the electron charge,
$\theta=(2m+1)\pi$ for
the $\nu=1/(2m+1)$ state and the term $\lambda |\phi |^4$ models the Coulomb
interaction which is present in the underlying microscopic theory. We choose
$A_0=0$. With the Bose field decomposed in terms of density and phase,
\be
\phi = \sqrt{\rho} \; e^{i\theta},
\ee
the field equations are
\be
\begin{array}{c}
\kappa\nabla^2\sqrt{\rho}
     = \left[ a_0 + V - \mu + \p_t\theta 
     + \kappa \left({\cal A} - \nabla\theta \right)^2 + 2\lambda\rho \right]
\sqrt{\rho}
                                                            \label{eom1} \\
\p_t \rho + \nabla\cdot{\bf j} = 0   \label{cont1}
\end{array}
\ee
and
\be
\begin{array}{c}

\nabla\times{\bf {\cal A}} = -2\theta\left( \rho - \rhob \right)
\label{CSconstr}\\
\p_x a_0 + \p_t a_x
     = -2\theta j_y \\
\p_y a_0 + \p_t a_y
     = 2\theta j_x ,\\
                                                           \label{cont}
\end{array}
\ee
where ${\bf {\cal A}} \equiv {\bf a}+e{\bf A}$, $\rhob\equiv eB/2\theta$, 
and the current density is
\be
{\bf j} = -2\kappa\rho\left( {\bf \cal A} - \nabla\theta \right).
\ee
The Hamiltonian is
\be
H=\int d^2x\left[ \kappa |{\bf D}\phi|^2 + V|\phi|^2 + \lambda |\phi|^4
 -  \mu |\phi|^2 \right],
                                                  \label{Ham}
\ee
where ${\bf D}=\nabla - i {\bf \cal A}$ and the constraint (\ref{CSconstr})
is implicit.

In some parts of the paper it will be convenient to work with dimensionless
quantities.
Rescaling the fields in (\ref{model}) such that length is measured in units
of the magnetic length $l_B=1/\sqrt{eB}$,  chemical potential in 
units of the cyclotron
energy $\omega_c=2e\kappa B$, time in units of the inverse cyclotron 
frequency and density in units of its constant bulk
value,
\be
\begin{array}{c}
{\tilde a}_0=a_0/\omega_c,~~~{\tilde\mu}=\mu/\omega_c      \nonumber\\
{\tilde\p}_t = \p_t/\omega_c                               \nonumber\\
{\tilde \p}_i = l_B \p_i,~~~{\tilde a}_i = l_B a_i          \label{rescale} \\
{\tilde \rho} = \rho/{\bar \rho} ,                           \nonumber
\end{array}
\ee
one finds that the Lagrangian (rescaled by an overall factor
$2\theta/2\kappa e^2 B^2$)
becomes
\be
{\cal L} = \phit^*\left( i{\tilde\p}_t - \at_0 - {\tilde V} 
         + \mut \right) \phit
         - \frac{1}{2} \left| ({\tilde\nabla} - i{\tilde {\cal A}})\phit
           \right|^2
         - \lt \left| \phit \right|^4
         + \frac{1}{2} \e^{\mu\nu\lambda}\at_{\mu}
                          {\tilde \p}_{\nu}\at_{\lambda}
                                         \label{dimlessL}
\ee
with
\be
\lt\equiv \frac{\lambda}{4\kappa\theta}.                     \label{Adef}
\ee
Thus, the theory contains only one free, dimensionless parameter (in
addition to the
chemical potential) \cite{tafelmayer1}. In particular it is worth noting
that the
statistics parameter $\theta $ is absorbed in the rescaled interaction
parameter $\tilde
\lambda$.


\section{Bulk properties}     \label{bulksec}

We start by reviewing the main features of the CSGL model in the absence of an
external potential, $V({\bf x})=0$. It is known for the FQHE that at
special fillings,
$\nu=1/(2m+1)$, the ground state is characterized by a {\em uniform}
electron density,
${\bar \rho} = \nu eB/2\pi$. As is well known \cite{zhang1}, the CSGL model
has this
constant solution when the chemical potential takes a special value $\mu_0$,
\be
    \rho &=& {\bar \rho} 
          = \frac{\mu_0}{2\lambda} 
          = \frac{eB}{2\theta}
                                                           \label{MFGS} \\
 {\bf a} &=& -e{\bf A} \\
     a_0 &=& 0.
\ee
This solution simultaneously minimizes the kinetic term ($D_i\phi = 0$) and the
potential $V(\phi) = \mu|\phi |^2 - \lambda |\phi |^4$.

However, there exist other static solutions with the same constant density
$\rhob$ away
from the potential minimum. These solutions have $a_0\neq 0$ and
$\mu\neq\mu_0$ with
$\mu-a_0=\mu_0$. This is because the chemical potential enters the
equations of motion
only in the combination $\mu-a_0$; thus, a shift in the chemical potential
can always
be absorbed as a constant shift in $a_0$. In this sense, the solutions of
the equations
of motion are independent of the value of $\mu$. However, the energy {\em
does} depend
on $\mu$ as can be seen from Eq.(\ref{Ham}). This observation is useful for
understanding how the QH plateaux emerge in this model, as we will discuss
below (see
also Curnoe and Weiss \cite{curnoe1}).

The low energy charged excitations of the QH system are the localized {\em
quasiparticles} and {\em quasiholes}. These are believed to be formed when
the density
is changed relative to the special filling $\rho=\bar \rho$, {\em e.g.} by
changing the
strength of the magnetic field $B$. In the CSGL model these excitations are
represented by vortex and anti-vortex solutions, whose existence has been
shown in
\cite{zhang1,zhang2} and their properties examined by several authors
\cite{ezawa1,tafelmayer1,curnoe1}. (We refer to the solutions with
charge deficit as the vortex solutions and the solutions with excess charge as
the anti-vortex solutions.) A general vortex solution is characterized by a
topological
quantum number $s$, which is $s=\pm1$ for the elementary vortex/anti-vortex
solutions.
The corresponding charges are $\pm \nu e$, and the energies (relative to
the ground
state) we denote by ${\e}^v$ for the vortex and ${\e}^{av}$ for 
the anti-vortex.

For given $s$ and $\tilde\mu$, the dimensionless vortex/anti-vortex 
energies are
functions of $\lt$ only. These functions were studied numerically by
Tafelmayer in Ref.\,\cite{tafelmayer1}, for $s=\pm 1,\pm2$ and ${\tilde
\mu}={\tilde \mu}_0$. As demonstrated by these functions there are two
separate regions
of $\lt$ where the vortex solutions show qualitatively different behaviour.
The regions
are separated by the special point $\lt = 1/2$, referred to as the {\em
self-dual point}. At this value of $\lt$ the vortices are solutions of a linear
(self-dual) differential equation, and the vortex energy is exactly
proportional to the
vorticity. Thus, the vortices are non-interacting. When $\lt>1/2$, the
energy of $n$
vortices with topological charge $1$ is {\em lower} than that of 
one vortex with
vorticity
$n$ (as demonstrated in Ref.\,\cite{tafelmayer1} for $n=2$). This means that a
stable multi-vortex configuration with $s=1$ can exist, since the vortices {\em
repel} each other. For
$\lt<1/2$ the situation is opposite, so vortices {\em attract} each other;
$n$ vortices
with $s=1$ would tend to combine to one large vortex. We may then identify
the interval
$\lt\geq 1/2$ as physically relevant for the FQH effect, since only in this
interval a
stable density of vortices will be formed when the filling 
fraction is decreased
relative to the plateau value
$1/(2m+1)$.
For anti-vortices there is no self-dual point. They have a repulsive
interaction for
all values of $\lt$.

The existence of finite-energy vortex/anti-vortex solutions together with
the freedom to
vary the chemical potential for a given solution leads to the
incompressibility of the
ground state and the existence of QH plateaux within the CSGL model. We
argued above
that a given solution of the equations of motion is independent of the
value of $\mu$,
whereas the energy of the solution {\em does} depend on $\mu$. The energy
(\ref{Ham})
can be rewritten in a dimensionless form as
\be
{\tilde E}({\tilde {\mu}}) = {\tilde E_0} - {\tilde {\mu}{\tilde Q}},
\ee
where ${\tilde Q}=2\theta Q =2\theta\int d^2x\;\rho$, 
${\tilde{\mu}}=\mu/2e\kappa B$,
${\tilde E}=2\theta E/2e\kappa B$, and ${\tilde E_0}$ is independent of 
${\tilde{\mu}}$. (The occurrence of the additional (dimensionless) factor
$2\theta$ in the rescaling of the {\em total} energy and charge is due 
to our choice to include this factor in the rescaling of the density.) 
For ${\tilde\mu}={\tilde\mu}_0$ the energy is minimized by the
constant solution ${\tilde\rho}=1$, and the vortex- and anti-vortex
solutions have
positive energies, ${\tilde {\e}}^v({\tilde {\mu}_0})$  and 
${\tilde {\e}}^{av}({\tilde
{\mu}_0})$ respectively, relative to the ground state. When the
chemical potential is increased by $\delta {\tilde {\mu}}$, the energies of
all the
solutions will decrease by  $\delta {\tilde {\mu}}$ times their respective
charge.
Since the anti-vortex configuration has an excess charge as compared to
${\tilde\rho}=1$, its energy will be lower than that of the uniform
solution when
${\tilde {\mu}}$ exceeds a critical value
\be
\tilde {\mu}^+ = \tilde {\mu}_0+{\tilde {\e}}^{av}({\tilde {\mu}_0})/2\pi
\ee
with $2\pi$ as the (rescaled) charge of the anti-vortex. Thus, for ${\tilde
\mu} \gtrsim
\tilde {\mu}^+$ the uniform ground state becomes unstable with respect to
anti-vortex
formation.
Similarly, a lower critical value exists,
\be
\tilde {\mu}^- = \tilde {\mu}_0-{\tilde {\e}}^{v}({\tilde {\mu}_0})/2\pi,
\ee
and if this is exceeded the uniform ground state becomes unstable with
respect to
vortex formation. Thus, there is a total ``window'' in the (dimensionless)
chemical
potential,
\be
\Delta {\tilde {\mu}} =
       \left({\tilde {\e}}^{v}({\tilde {\mu}_0}) + {\tilde
{\e}}^{av}({\tilde {\mu}_0})
                          \right)/2\pi
\ee
within which the ground state is given by the uniform density solution.

Since ${\tilde \mu}=\mu/2e\kappa B$, a variation in $\tilde \mu$ can be
interpreted
either as a variation in $\mu$ at fixed $B$,
\be
\delta {\tilde\mu} = \frac{1}{2e\kappa B} \delta\mu
\ee
or as a variation in $B$ at fixed $\mu$,
\be
\delta {\tilde {\mu}} = -\frac{\mu}{2e\kappa B^2} \delta B,
\ee
corresponding to the following interpretations (returning to dimensionful
quantities):

\noindent
{\bf 1)} The ground state density at fixed $B$ remains unchanged 
within a finite
interval
\be
\Delta\mu = \left( \e^v(\mu_0) + \e^{av}(\mu_0) \right)/\nu  \label{incompr}
\ee
which implies the incompressibility condition
\be
\left( \frac{\p\rho}{\p\mu} \right)_B = 0
\ee
near $\mu=\mu_0$.\\
\noindent
{\bf 2)} The density remains ``locked'' to the magnetic field through the
condition $b+eB=0$ when $B$ is varied within a finite interval
\be
\Delta B = \left( \e^v(\mu_0) + \e^{av}(\mu_0) \right) 
           \frac{\pi}{e\nu^2\lambda}
         = \left({\tilde \e}^v(\mut_0) + {\tilde \e}^{av}(\mut_0) \right)
           \frac{B}{4\pi\lt}.
                                                 \label{Delta B}
\ee
Since the Hall conductivity is proportional to $\rho/B$ this implies that
$\sigma_H$
is constant within the interval $\Delta B$, giving rise to the Hall plateaux.

So, at least {\em qualitatively}, the simple mean field theory reproduces
the right
picture as far as the plateaux are concerned. This is the case even without an
impurity potential, since this is a model with a fixed chemical 
potential rather
than fixed charge density. Thus, on a plateau the charge density changes with
the magnetic field until it is energetically favoured to create vortices or
anti-vortices in order to re-adjust the mean charge density. At this point
the Hall
conductivity leaves the plateau.

When the strength of $\lt$ is increased, the width of the plateaux will
decrease. This
follows from the expresssion (\ref{Delta B}) and the fact that 
the dimensionless
vortex-and anti-vortex energies divided by $\lt$ decrease with $\lt$ (see
Ref.\,\cite{tafelmayer1}). This is in accordance with the expectation that an
increase in the
(Coulomb) interaction of the particles will tend to suppress variations in
the charge
density. If an impurity potential is added we expect a broadening of the
plateaux beyond
the point where vortices are formed, since the vortices will bind to the
impurities and
therefore not affect the Hall conductance. This gives the connection to the
standard
picture of the fractional quantum Hall effect.


\section{The edge}        \label{edgesec}
In the presence of a confining potential $V({\bf x})$, corresponding to an
external
electric field $e{\bf E} = -\nabla V$, the size of the system becomes
finite, with a
density profile at the edge that depends on the form of $V$. We choose here
a geometry
such that $V$ is a function of $x$ only and thus translationally invariant
in the
$y$-direction. This means that the edge is parallel to the $y$-axis. We
further choose
$x=0$ to be an interior line in the bulk (far away from the edge), and
assume the
system to be symmetric about this line. Due to this symmetry it is 
sufficient to
consider the system for $x>0$. We first examine the ground state of the
system and
then, in Sec.\,\ref{eesec}, discuss the gapless edge modes obtained by
perturbing the
ground state. Static solutions have previously been obtained numerically by
Orgad and
Levit \cite{orgad1} in the case of an infinitely high step function
potential. In this
paper, we consider a smooth potential $V(x)$ which is more relevant for real
physical situations. \footnote{However, the translationally invariant form
we assume
for the ground state density at the edge may {\em implicitly} mean that
the steepness
of the potential should not be too small. A slowly varying potential may be
considered
as an
$x$-dependent chemical potential, and from the discussion of the bulk
properties, we
in this case expect a broad edge with vortices trapped in the edge region. This
corresponds in the FQHE to a composite edge with strips of compressible and
incompressible Hall fluids, as discussed in Ref.\,\cite{chamon1}.}


\subsection{Ground state}        \label{stsec}

We consider ground states which are translationally invariant in the
$y$-direction,
{\em i.e.} they correspond to static solutions of the field equations where
all the
fields are functions of $x$ only. The field equations then reduce to the form,
\be
\kappa\p_x^2\phi(x) &=& \left(a_0(x) - \mu + V(x) + \kappa {\cal A}_y(x)^2
                     + 2\lambda\rho\right)\phi(x)
                       \label{s1}\\
\p_x {\cal A}_y(x) &=& -2\theta \left( \rho(x) - {\rhob} \right)  \label{s2} \\
\p_x a_0(x) &=& 4\kappa\theta {\cal A}_y(x) \rho(x),         \label{s3}
\ee
where we have made the gauge choice $\phi=\sqrt{\rho}$, {\em i.e.} the
wave functions are real. Again, ${\cal A}_y = eA_y+a_y$ and due to the
symmetry of
the system (and our gauge choice) $a_x = 0$. As before, $A_x=0$ 
and $A_y=Bx$. We
further choose $a_y(0)=0$, and as follows from the discussion in the
previous section
$a_0(0)=\mu-\mu_0$ in order for the density to take the constant value
$\rhob$ in the
bulk.

There is a conserved charge in this problem, $Q=\int dx \rho$, and
consequently a
one-parameter set of ground states parametrized by this charge
\cite{orgad1}. This set
of ground states can alternatively be parametrized by the chemical
potential. Thus,
the ground state energy, which has the $\mu$-dependence $E=E_0(Q)-\mu Q$, may
initially be considered as depending on two independent parameters $Q$ and
$\mu$.
However, for the true ground state, which is found by minimizing the energy
with respect
to $Q$,
\be
\frac{\p E(\mu,Q)}{\p Q} = 0,
\label{dEdQ}\ee
the chemical potential will be a function of $Q$,
\be
\mu = \frac{d E_0}{d Q}.
\ee

From the discussion of the previous section we know that there is no change
in the bulk
charge for small variations in $\mu$ around $\mu_0$. This means that when
the chemical
potential is changed within the limits determined by the vortex- 
and anti-vortex
energies, all changes in the ground state take place at the edge. In
particular, the
additional charge which is introduced by a change in the chemical potential, is
confined to the edge.

The energy is stationary with respect to variations in the wave function about
the ground state. This is true, due to (\ref{dEdQ}), also for variations where
the total charge is changed. For an infinitesimal translation $\delta x$ of
the edge,
with $\delta Q = \rhob\delta x$ and $\delta\rho(x)=-\rho'(x)\delta x$, one finds
\be
\delta E_0 = \delta x \left( \lambda\rhob^2 - \int dx \; V(x)\rho'(x) \right)
\ee
which leads to
\be
\mu = \frac{\mu_0}{2} + \frac{1}{\rhob} \int dx\; V'(x)\rho(x).  \label{mu(V)}
\ee
The same expression, but with the additional term $a_0(\infty)$ on the
right hand side
can be derived from the equation of motion (\ref{s1}) by multiplying it
with $\phi'(x)$
and integrating over $x$. This leads to the conclusion $a_0(\infty)=0$. An
interesting consequence of this is that the total current in the system is
zero when
$\mu=\mu_0$: Integrating Eq.(\ref{s3}) over $x$ gives
\be
a_0(\infty) - a_0(0) = -2\theta J,
\ee
where $J$ is the total current (in the $y$-direction). Since
$a_0(0)=\mu-\mu_0$ we have
\be
\mu-\mu_0 = 2\theta J.                               \label{muJ}
\ee
Thus, the integrated current, which in general will have two contributions
of opposite
sign at the edge, will have an overall sign which changes at $\mu=\mu_0$.
\begin{figure}[htb] 
\begin{center}
\mbox{\psfig{figure=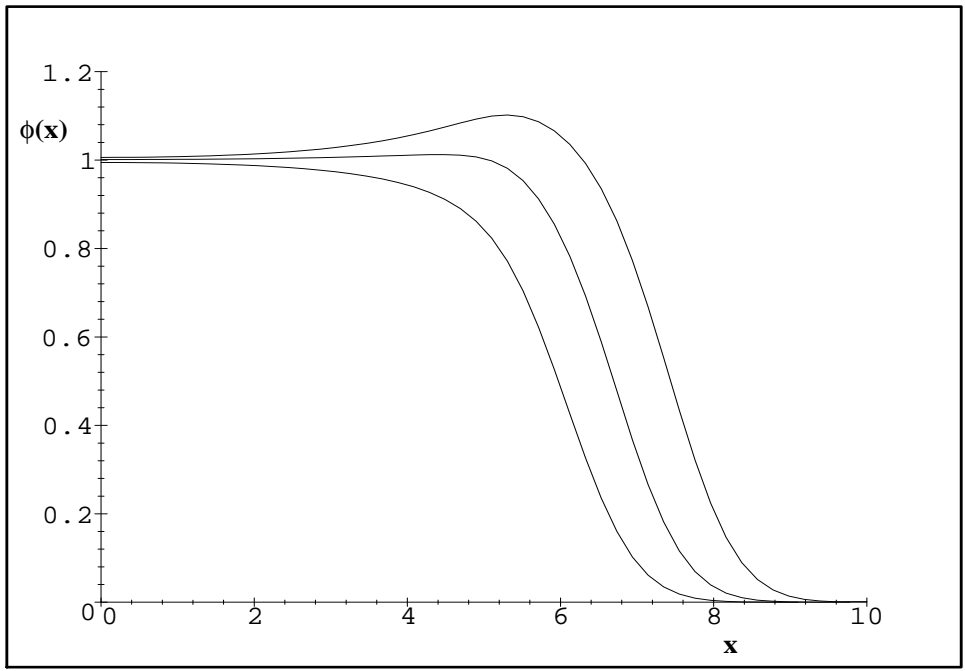,width=14cm,angle=270,height=10cm}}
\end{center}
\caption[]{\footnotesize Ground state edge profiles for $\lt =1$ and
${\tilde V}(x) = \frac{1}{2}\;\theta(x-5)\cdot (x-5)^2$.
The curves correspond to (from above) the upper critical value
$\mu^+$, $\mu=\mu_0$ and the lower critical value
$\mu^-$.}
\label{fig1}
\end{figure}

This change in sign of the current has an analogy in the microscopic
picture of the
real quantum Hall system: There are two contributions to the current. One
is the drift
current, caused by the drift of the cyclotron orbits in the external
electric field. The
other is the polarization current, due to the gradient of the density at
the edge and
with origin in the cyclotron motion of the electrons. These contributions
will have
opposite sign, and which one will dominate is determined by the value of
the chemical
potential or, equivalently, by the charge collected at the edge.

\begin{figure}[htb] 
\begin{center}
\mbox{\psfig{figure=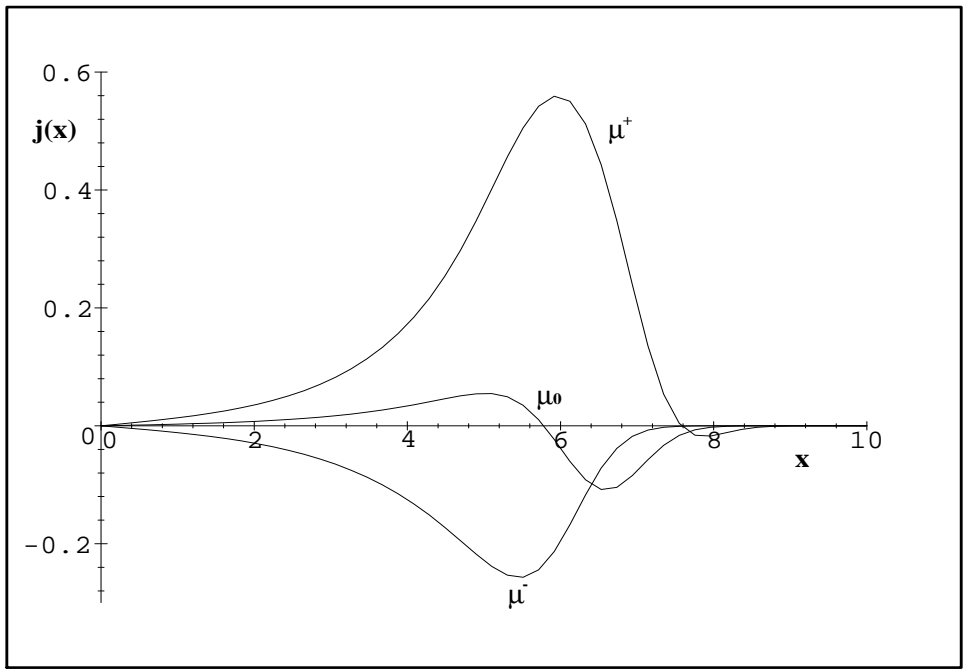,width=14cm,angle=270,height=10cm}}
\end{center}
\caption[]{\footnotesize  Current densities corresponding to the profiles in
Fig.\,\ref{fig1}. The total (static) current is negative at $\mu<\mu_0$, zero
at $\mu=\mu_0$ and positive for $\mu>\mu_0$.}
\label{fig2}
\end{figure}

A numerical study of charge and current densities in the ground state is
presented in
Figs.\,\ref{fig1} and \ref{fig2}.  
In Fig.\,\ref{fig1} the edge profile $\phi(x)$ is shown for three
different
values of the chemical potential in the case of a harmonic confining
potential ${\tilde V}(x) =
0.5\;
\theta(x-5)\cdot (x-5)^2$ where $\theta(x)$ is the step function, and
$\lt=\lambda/(4\kappa\theta)=1$. The upper curve corresponds to
$\mu=3.5\;\omega_c$
which is the upper critical value as estimated from the numerical values in
Ref.\,\cite{tafelmayer1}. Note the excess charge accumulated at the edge. For
$\mu$ larger
than this value the ground state is expected to be an anti-vortex
configuration in the
bulk rather than a state with uniform density. The middle curve corresponds to
$\mu=\mu_0=2\;\omega_c$ whereas $\mu=1.3\;\omega_c$, the estimated lower
critical
value, for the lowest curve. For smaller values of the chemical potential
the ground
state is believed to reorganize into a density of vortices.

The current density profiles corresponding to these three solutions are
shown in Fig.\,\ref{fig2}.
Note the change in direction of the total static current: For
$\mu=1.3\;\omega_c < \mu_0$ the current is negative; 
for $\mu=\mu_0$ the two contributions cancel (as
anticipated in Eq.(\ref{muJ})), whereas the positive contribution 
dominates when
$\mu>\mu_0$ as in the last plot.


\subsection{Edge excitations}                                   \label{eesec}

The linearized field equations for fluctuations of the fields about the
ground state have the form
\be
\kappa \nabla^2 \delta \sqrt{\rho} &=&
   \left[ \delta a_0 + \p_t\delta\theta
      + 2\kappa {\cal A}_y \left( \delta a_y - \p_y\delta\theta \right)
      + 2\lambda \delta\rho\right] \sqrt{\rho}  \nonumber \\
     &+& \left[ a_0 + V - \mu + \kappa {\cal A}^2 + 2\lambda\rho \right]
                                           \delta \sqrt{\rho}
                                                               \label{RPA1}\\
\nabla\times{\bf \delta a} &=& -2\theta\delta\rho              \label{RPA2}\\
\p_x\delta a_0 + \p_t\delta a_x
     &=& 4\kappa\theta \left[ {\cal A}_y\delta\rho
      + \rho\left(\delta a_y - \p_y\delta\theta  \right) \right]
                                                               \label{RPA3}\\
\p_y\delta a_0 + \p_t \delta a_y
     &=& -4\kappa\theta\rho\left( \delta a_x - \p_x\delta\theta \right)
                                                               \label{RPA4}\\
\p_t\delta\rho
     &=& 2\kappa\nabla\cdot\left[ {\bf {\cal A}}\delta\rho
      + \rho \left( \delta{\bf a} - \nabla\delta\theta \right) \right].
                                                               \label{RPA5}
\ee
For low-frequency fluctuations propagating along the edge we write, as a simple
{\em ansatz}, all the fields as an
$x$-dependent part modulated by a slowly varying ($y,t$)-dependent 
harmonic part,
\be
\delta\rho(x,y,t)     &=& \delta\rho(x)\; e^{i(qy-\omega t)}   \nonumber \\
\delta\theta(x,y,t)   &=& -i\eta(x)\; e^{i(qy-\omega t)}     \nonumber   \\
\delta a_{\mu}(x,y,t) &=& \delta a_{\mu}(x)\; e^{i(qy-\omega t)}.
                                                        \label{modfields}
\ee
The field equations can then be reduced to purely $x$-dependent equations
with the prescription $\p_y\rightarrow iq$ and $\p_t\rightarrow -i\omega$.
We consider
$q$ as a small parameter and expand all the fields in powers of
$q$, {\em e.g.}
\be
\delta\rho(x) = \delta\rho^{(0)}(x) + q\;\delta\rho^{(1)}(x)
              + q^2\;\delta\rho^{(2)}(x) + \cdots \; .            \label{qexp}
\ee
Since we are interested in the gapless modes we assume $\omega$ to be of first
order in $q$. The field equations (\ref{RPA1})--(\ref{RPA5}) are examined
order by order
in $q$. As a starting point we assume that to lowest order only the phase
fluctuation $\eta$ contributes.  This reduces the $0^{th}$ order equations to
\be
4\kappa\theta(\rho\p_x\eta) = \p_x\left( \rho\p_x\eta \right) = 0 \label{eta0}
\ee
so $\eta^{(0)}$ is a constant.

To first order in $q$, in (\ref{RPA4}) and (\ref{RPA5}) only the
terms involving $\rho(\delta a_x - \p_x \theta) \propto \delta j_x$
survive, simply
giving
\be
\delta j_x^{(1)} = 0.
\ee
The first three equations (\ref{RPA1})--(\ref{RPA3}), on the other hand,
take the same
form as those for a first order variation in the equations of
motion (\ref{s1})--(\ref{s3}) for the $Q$-dependent ground states.
Thus, to
${\cal O}(q)$ we have the following solution:
\be
\delta\rho^{(1)}(x) &=& \delta\rho^{(s)}(x) \\
\delta a_0^{(1)}(x) &=& \delta a_0^{(s)}(x) - \delta\mu 
                     + \frac{\omega}{q}\eta^{(0)} \\
\delta a_y^{(1)}(x) &=& \delta a_y^{(s)}(x) + \eta^{(0)}
\ee
where $^{(s)}$ refers to variations between stationary fields obtained by a
small change
$\delta Q$ in the ground state charge and a corresponding change $\delta
\mu$ in the
chemical potential.

For the stationary ground state fields the current conservation is trivially
satisfied. This is not the case for the modulated fields (\ref{modfields}),
where the
current conservation in fact determines the dispersion law for the edge
waves. We note
that to lowest order the current conservation involves the second order
contribution to
$j_x$ and can be used to determine this from the first order fields.
However, the
integrated equation contains only first order fields, due to the boundary
condition
\be
j_x(0)=j_x(\infty)=0
\ee
and has the form
\be
\omega \delta Q=q \delta J.
\ee
This equation is in fact correct to all orders in $q$, with $\delta Q$ as the
fluctuation in the integrated charge and $\delta J$ as the fluctuation of
the total
current in the $y$-direction. To first order in $q$ this gives
\be
\frac{\omega}{q}=\frac{dJ}{dQ}=\frac{1}{2\theta}\,\frac{d\mu}{dQ}
\label{disp1}
\ee
where the derivatives refer to the $Q$-dependent ground state. This
expression for
the dispersion law agrees with the expression found in Ref.\,\cite{orgad1} in
the case of
an infinitely steep confining potential.  In our case we can further 
express the
dispersion in terms of the potential $V(x)$. Inserting the expression
(\ref{mu(V)})
for $\mu$ derived in the previous section into (\ref{disp1}) gives
\be
\frac{\omega}{q} = \frac{1}{eB} \int_0^{\infty} dx\; V'(x) 
                  \frac{d\rho}{d Q}.
\label{Vdisp}
\ee
For a confining potential which is approximately linear over the edge it
has the form
\be
\frac{\omega}{q} = -\frac{E}{B},
\ee
with $eE=-V'$, and more generally it can be read as an averaged form of
this equation,
with the $E$-field averaged over the edge. We note that the velocity 
of the edge
waves is always in the same direction as the drift velocity of the
electrons, and
is not necessarily in the same direction as the (integrated) edge current
in the ground
state.


\section{Two-component system}        \label{twocompsec}

\subsection{The model}

The CSGL Lagrangian (\ref{model}) can be extended to include spin
\cite{lee1}. This is
done by replacing $\phi$ by a two-component complex scalar field and
introducing a
Zeeman term
\be
{\cal L}_{Zeem} = -\frac{g\mu_B B}{2}\phi^{\dagger}\sigma_z \phi.
\ee
The model then takes the form
\be
{\cal L} &=& \sum_{\alpha=1}^2 \phi_{\alpha}^* \
                    \left( i\p_t - a_0 + \mu_{\alpha} -V \right)\phi_{\alpha}
         - \kappa\sum_{\alpha=1}^2 \left| (\nabla - i{\cal A})
           \phi_{\alpha}\right|^2
                                               \nonumber \\
        &-& \lambda (\rho_1 + \rho_2)^2
         + \frac{1}{4\theta}\e^{\mu\nu\lambda}a_{\mu}\p_{\nu}a_{\lambda},
                                                           \label{model2}
\ee
where the index $\alpha$ denotes the two spin states (up and down), so that
$\rho=\rho_1+\rho_2$ is the total (charge) density, while
$\rho_S=\rho_1-\rho_2$ is the
spin density. A spin unit vector may be introduced by
\be
{\bf n} = \frac{1}{\rho}\phi^{\dagger} {\bf {\sigma}} \phi
\ee
with components in the plane as well as in the direction of the $B$-field
($z$-direction). In (\ref{model2}) the Zeeman term is represented by the
difference
between the chemical potentials of the two components.

With the densities and phases of the two components of $\phi_\alpha$ as
variables,
the field equations are,
\be
\begin{array}{c}
\kappa\nabla^2\sqrt{\rho_{\alpha}} =
      \left[ a_0 + V - \left( \mu_{\alpha} - \p_t \theta_{\alpha}  \right)
    + \kappa \left( {\bf {\cal A}} - \nabla\theta_{\alpha} \right)^2
    + 2\lambda \left( \rho_1 + \rho_2 \right) \right] \sqrt{\rho_{\alpha}}\\
\nabla \times {\bf a} =
    - 2\theta \left( \rho_1 + \rho_2 \right) \\
\p_x a_0 + \p_t a_x =
      4\kappa\theta \sum_{\alpha} \rho_{\alpha}
      \left( {\cal A}_y - \p_y \theta_{\alpha} \right)       \label{2RPA} \\
\p_y a_0 + \p_t a_y =
    - 4\kappa\theta \sum_{\alpha} \rho_{\alpha}
      \left( a_x - \p_x \theta_{\alpha} \right) \\
\p_t \rho_{\alpha} + \nabla\cdot {\bf j}_{\alpha} = 0,
\end{array}
\ee
where ${\cal A}_i = a_i + eA_i$ as before, and the current densities are
given by
\be
{\bf j}_{\alpha}
      = - 2\kappa \rho_{\alpha} \left( {\bf {\cal A}} - \nabla
\theta_{\alpha} \right).
\ee
This system has two conserved charges
$Q_{\alpha}\equiv\int d^2x\; \rho_{\alpha}$ .

\subsection{Ground state properties}

The form of the Hamiltonian
\be
{\cal H}=\kappa\sum_{\alpha=1}^2 \left| (\nabla - i{\cal A})\phi_{\alpha}
\right|^2
        + V(\rho_1+\rho_2) + \lambda(\rho_1+\rho_2)^2 
        -\mu_1\rho_1 - \mu_2\rho_2
\ee
implies that the bulk ground state ($V=0$) is fully polarized and has a 
uniform
density,
\be
\rho_1 &=& \rhob = \frac{eB}{2\theta} \equiv\frac{\mu_0}{2\lambda} \\
\rho_2 &=& 0,
\ee
where we have assumed $\mu_1>\mu_2$. The chemical potential of the upper
component is
related to $\mu_0$ by
\be
\mu_1-a_0(0)=\mu_0.
\ee

\bigskip
\noindent
Even though the ground state in the bulk is fully polarized, this may not
be the case
at the edge of the system. Tilting the spins ({\em i.e.} $\rho_2\neq 0$) at
the edge
may lower the energy if the Zeeman gap is not too large. This problem was
recently
studied numerically in Ref.\,\cite{karlhede1}; their calculations were based
on Hartree-Fock and an effective action approach \cite{sondhi1} in the
low-energy limit
of the CSGL model. Here, we shall examine some of the ground state
properties at the
edge both analytically and numerically, within the framework of CSGL theory.

As before, we introduce a confining potential $V(x)$, which gives rise to a
straight
edge parallel to the $y$-axis. We consider stationary solutions, with 
densities
$\rho_1$ and $\rho_2$ that are translationally invariant in the
$y$-direction. With the
gauge choice $\theta_1=0$, the solutions take the general form
\be
\phi_1 &=& \sqrt{\rho_1(x)}                            \label{stat1}\\
\phi_2 &=& \sqrt{\rho_2(x)} e^{i(ky-\beta t)}          \label{stat2}
\ee
with $\rho_1=\rhob,\;\rho_2=0$ in the bulk. The gauge choice implies
$A_x=a_x=0$ with
$a_0$ and $a_y$ as functions of $x$ only. The parameters $k$ and $\beta$ are
undetermined at this stage. However, we see that if $k\neq 0$ this means
that the spins
will rotate around the $B$-field  when moving along the edge. With the
ansatz (\ref{stat1}) and (\ref{stat2}) the equations of motion reduce to
\be
\kappa\p_x^2 \sqrt{\rho_1} &=& \left[ a_0 + V - \mu_1 + \kappa {\cal A}_y^2
                            + 2\lambda(\rho_1+\rho_2) \right] \sqrt{\rho_1}
        \label{S1} \\
\kappa\p_x^2 \sqrt{\rho_2} &=& \left[ a_0 + V - \mu_2 - \beta 
                            + \kappa ({\cal A}_y-k)^2
                            + 2\lambda(\rho_1+\rho_2) \right] \sqrt{\rho_2}
        \label{S2} \\
\p_x {\cal A} &=& -2\theta \left( \rho_1 + \rho_2 - \rhob \right)
\label{S3} \\
\p_x a_0 &=& 4\kappa\theta \left[ \rho_1 {\cal A}_y + \rho_2 ({\cal A}_y - k)
             \right]
                                                                    \label{S4}
\ee
The corresponding energy (per unit
length in the
$y$-direction) is
\be
E &=& \int dx\; \left[ \kappa\left( \p_x \sqrt{\rho_1} \right)^2
  + \kappa \left( \p_x \sqrt{\rho_2} \right)^2
  + \kappa {\cal A}_y^2 \rho_1 + \kappa ({\cal A}_y-k)^2 \rho_2 
  + V (\rho_1 + \rho_2)
                                      \right.\nonumber \\
  &+& \left. \lambda (\rho_1 + \rho_2)^2 \right]
  - \mu_1 Q_1 - \mu_2 Q_2,                         \label{twocompen}
\ee
where $Q_{\alpha}$ now denotes charge per unit length,
$Q_{\alpha}=\int dx\;\rho_{\alpha}(x)$.

The ground state conditions are
\be
\frac{\p E}{\p Q_1} = \frac{\p E}{\p Q_2} = 0
\label{QGS}
\ee
and
\be
\frac{\p E}{\p k} = 0.      \label{kGS}
\ee
These conditions determine the chemical potentials as functions of 
the charges,
$\mu_{\alpha} = \mu_{\alpha}(Q_1,Q_2)$. So here we have a 
{\em two-parameter set} of
ground state solutions parametrized by the charges $Q_1$ and $Q_2$.

From the condition (\ref{QGS}), that the ground state energy is invariant
under any
first order variation of the two charges, we can derive expressions for
$\mu_1$ and
$\beta$. We first consider a perturbation in $Q_1$ caused by an
infinitesimal translation of both components in the $x$-direction (thus
keeping $Q_2$
fixed). This gives
\be
\mu_1 = \lambda\rhob + \frac{1}{\rhob} \int dx\; V'(x) \rho(x)
\label{2mu(V)}
\ee
which corresponds to Eq.(\ref{mu(V)}) in the one-component case. 
Again, the same
expression but with an additional term $a_0(\infty)$ can be derived
directly from the
equations of motion, giving $a_0(\infty)=0$ in the ground state, and
consequently
\be
\mu_1 - \mu_0 = 2\theta J,                     \label{mu1J}
\ee
where $J=J_1 + J_2$ is the total (integrated) current along the edge.

Similarly, we can slightly perturb the ``spin charge'' $Q_S = Q_1 - Q_2$
keeping the
total charge fixed. This is done by an infinitesimal $SU(2)$ 
rotation of $\phi$,
\be
\phi_1 &\lra& \phi_1 + \ve\phi_2 \\
\phi_2 &\lra& \phi_2 -\ve^* \phi_1
\ee
with
\be
\ve = \ve(y) = \ve_0 e^{-iky}
\ee
and $\ve_0$ as an infinitesimally small real constant.
This introduces the following variations in the charges,
\be
\delta Q_S = 4\ve_0 \int\; dx \sqrt{\rho_1(x)}\sqrt{\rho_2(x)},~~
\delta Q = 0,
\ee
and the condition of stationary energy, $\p E/\p Q_S = 0$, gives
\be
\mu_1 - \mu_2  = -\kappa k^2
    + 2\kappa k \frac{\int dx\; {\cal A}_y(x) \sqrt{\rho_1(x)}\sqrt{\rho_2(x)}}
      {\int dx\; \sqrt{\rho_1(x)}\sqrt{\rho_2(x)}}.
                                                    \label{mu1-mu2}
\ee
The same expression, but with an additional factor $\beta$ on the right
hand side, can
be deduced from the equations of motion by multiplying (\ref{S1}) and
(\ref{S2}) by
$\sqrt{\rho_2}$ and $\sqrt{\rho_1}$ respectively, subtracting the two
equations and
integrating over $x$. This leads us to the conclusion that the ground state
condition
$\p E/\p Q_S = 0$ implies $\beta = 0$. Thus, there is no $t$-dependence in
the phase of
$\phi_2$. Note that the relations (\ref{mu1J}) and
(\ref{mu1-mu2}) were derived only demanding that the energy be minimized
with respect
to the charges; they are therefore valid even for values of $k$ that do not
minimize
the energy.
Finally, also demanding $\p E/\p k = 0$ one finds
\be
J_2 = -2\kappa\int dx\; \rho_2(x) \left( {\cal A}_y(x)-k \right) = 0
\ee
which gives an expression for the ground state value of $k$,
\be
k \equiv k_0 = \frac{1}{Q_2} \int dx\; \rho_2(x) {\cal A}_y(x).
                                                        \label{globalk}
\ee

As already stated, we expect a critical value of the Zeeman gap 
$\Delta \mu = \mu_1-\mu_2$ to exist, 
so that if this critical value is exceeded the
system will be
fully polarized also at the edge. If the gap is close to, but smaller than
this value we
expect the second component of the wave function to be small. We will make an
analysis of this case and illustrate the case by numerical evaluations of
the ground state wave function.

\begin{figure}[htb] 
\begin{center}
\mbox{\psfig{figure=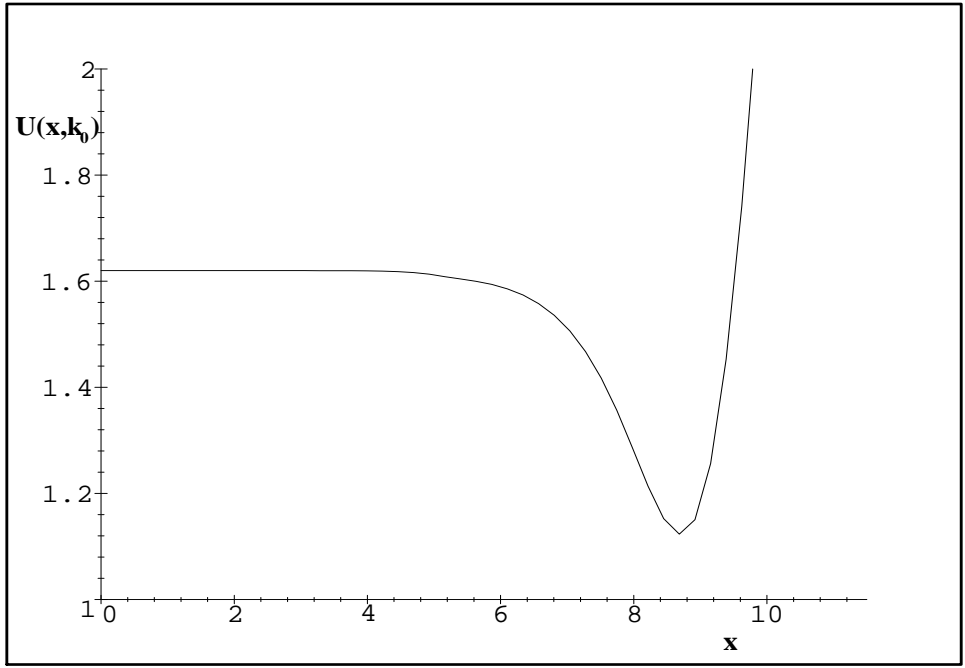,width=14cm,angle=270,height=10cm}}
\end{center}
\caption[]{\footnotesize  The (dimensionless) potential 
$U(x, {\tilde k}_0)$ determining the density
${\hat\rho}_2(x)$ of the lower component in the limit $\rho_2\ll\rho_1$. 
Here,
the confining potential is ${\tilde V}(x)=0.05\;\theta(x-5)\cdot (x-5)^2$,
$\mut_1\approx 1.5$ and ${\tilde k_0} = k_0\cdot l_B = 0.49$.}
\label{fig3}
\end{figure}
For small $Q_2$ we can expand $\mu_2$ in powers of this charge (for given
$Q_1$ and $k$)
\be
\mu_2 = \mu_2^{(0)} + \mu_2^{(1)} Q_2 + \cdots
\ee
and similarly for the density $\rho_2$,
\be
{\hat \rho}_2 \equiv \rho_2/Q_2
                  = {\hat \rho}_2^{(0)} + {\hat \rho}_2^{(1)} Q_2 \cdots.
\ee
The lowest order terms  correspond to the limit $Q_2\rightarrow 0$. This
means that
$\mu_2^{(0)}$ is the value of $\mu_2$ which corresponds to the critical
Zeeman gap for
given $\mu_1$. Similarly, ${\hat \rho}_2^{(0)}$ represents the profile of
$\rho_2$ in
the same limit. In this limit $\rho_2$ is negligible compared to $\rho_1$,
and the
problem is simplified since the equations for the two densities
((\ref{S1})--(\ref{S4})) then decouple: Eqns.(\ref{S1}), (\ref{S3}) and
(\ref{S4}) get
reduced to the single-component equations (\ref{s1})--(\ref{s3}) discussed
in 
Sec$.\,$\ref{stsec}, and the equation (\ref{S2}) for 
$\sqrt{\hat \rho_2}$ simply
becomes a
one-dimensional potential problem,
\be
\kappa\p_x^2\sqrt{\hat \rho_2(x)} = \left[ U(x,k) - \mu_2
\right]\sqrt{\hat \rho_2(x)}
                                                        \label{potpr}
\ee
where the potential is
\be
U(x,k) = a_0(x) + V(x) + \kappa ({\cal A}_y(x)-k)^2 + 2\lambda\rho_1(x).
\ee
\begin{figure}[htb] 
\begin{center}
\mbox{\psfig{figure=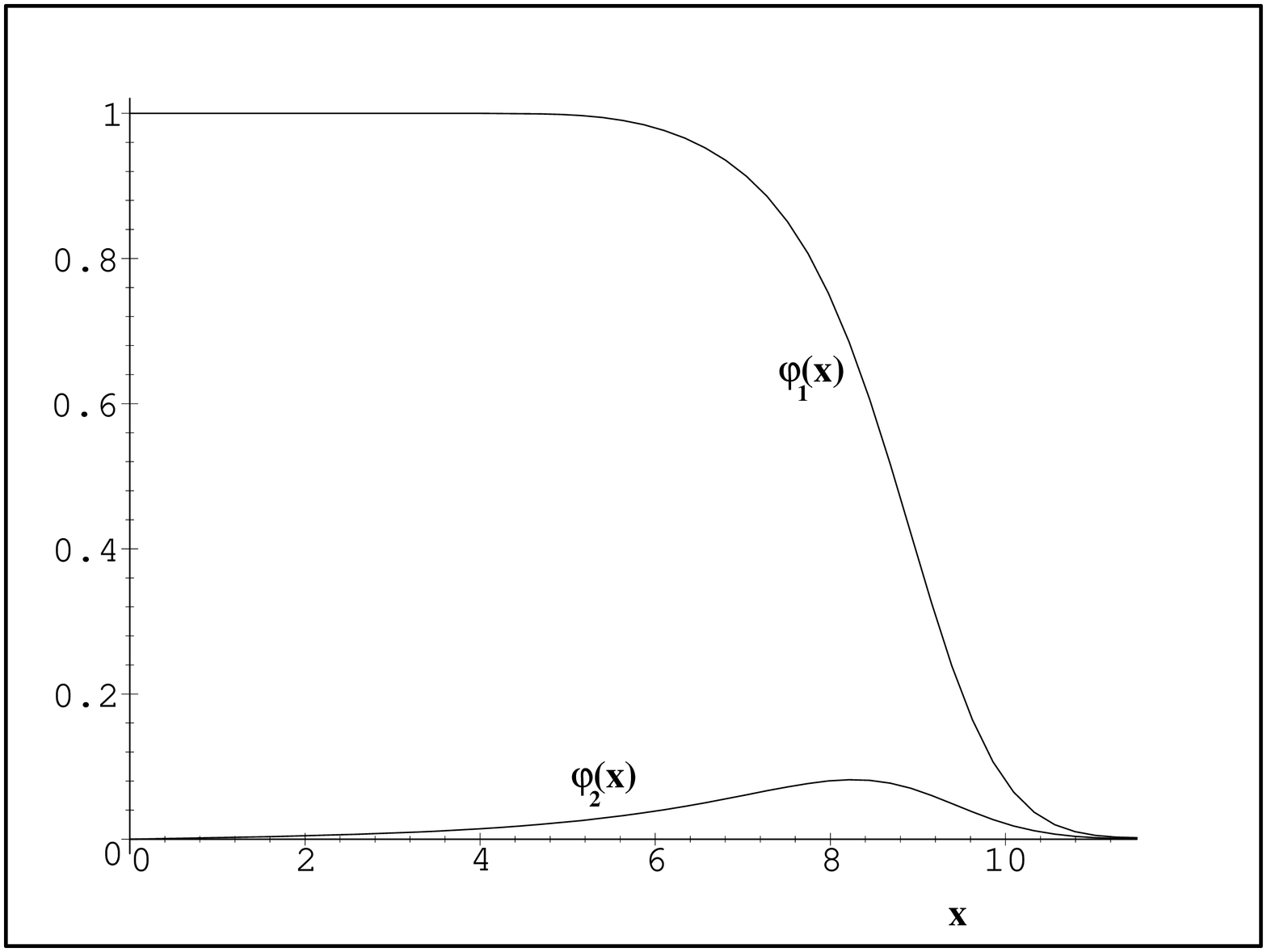,width=14cm,angle=270,height=10cm}}
\end{center}
\caption[]{\footnotesize Edge profiles 
$\sqrt{{\tilde \rho}_1(x)}\equiv\varphi_1(x)$ and 
$\sqrt{{\tilde \rho}_2(x)}\equiv\varphi_2(x)$ corresponding to
the potential in Fig.\,\ref{fig3}. 
The corresponding (critical) Zeeman gap is  $\mut_1-\mut_2\approx 0.002$.}  
\label{fig4}
\end{figure}

The ground state is found by minimizing the energy with respect to $k$, and
the chemical
potential $\mu_2$ is determined as the (lowest) eigenvalue of the potential
problem.
This determines the critical value of the Zeeman gap for the chosen value
of $\mu_1$.

The expansion in $Q_2$ may be used to correct iteratively the lowest order
results. Thus, the lowest order contribution to $\hat \rho_2$ gives first 
order
corrections to the equations for $\rho_1,\; a_0,\; {\cal A}_y$. 
This in turn gives
first order corrections to the potential $U(x,k)$ and to 
$\mu_2,\;\hat\rho_2$ and
$k_0$, etc. This gives a systematic way to determine the ground state
density profiles
for small
$Q_2$.

\begin{figure}[htb] 
\begin{center}
\mbox{\psfig{figure=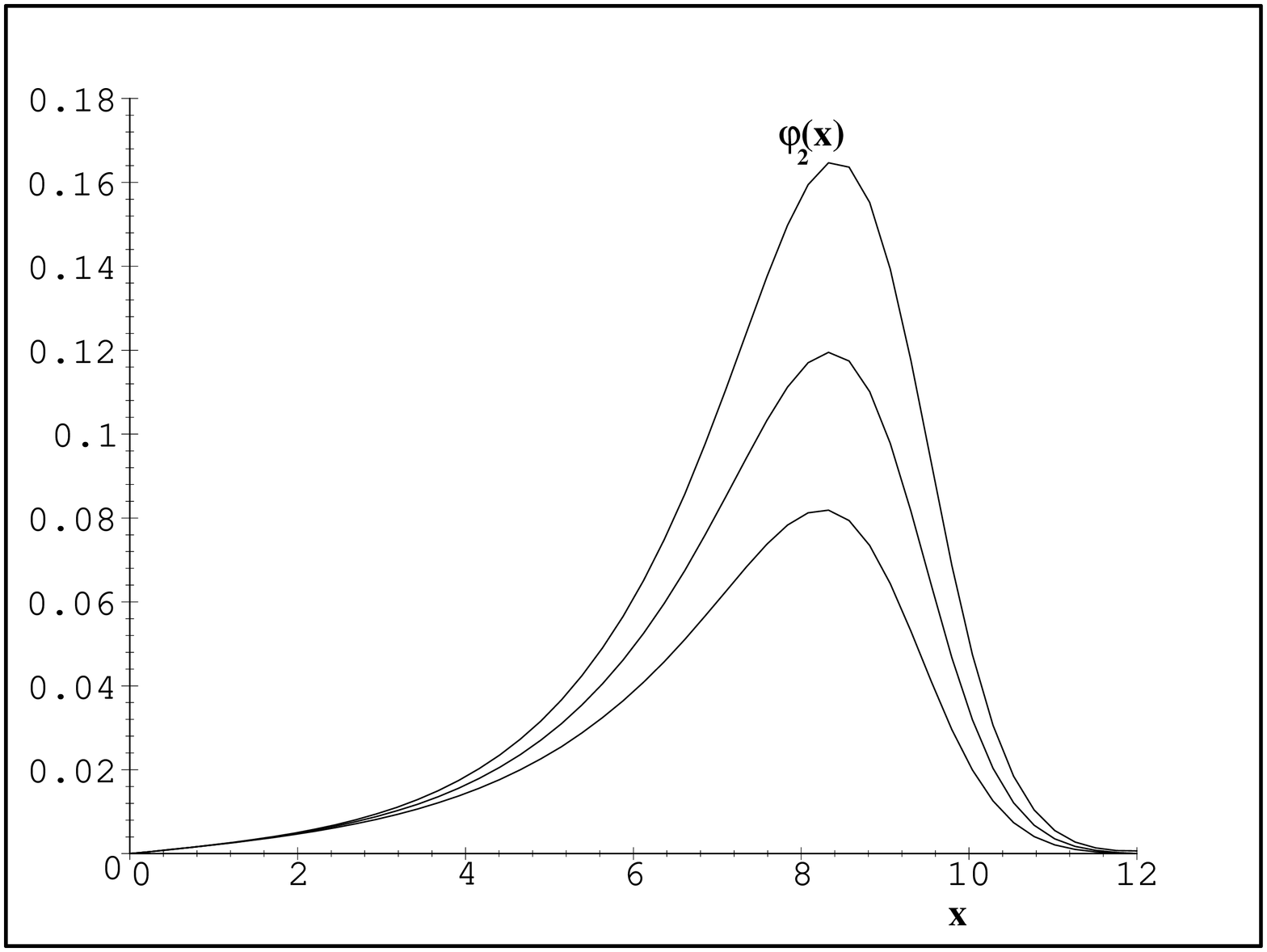,width=14cm,angle=270,height=10cm}}
\end{center}
\caption[]{\footnotesize Shape of $\varphi_2(x)=\sqrt{{\tilde \rho}_2(x)}$ 
for values of $k$ away from $k_0$.
The confining potential is 
${\tilde V}(x)=0.05\;\theta(x-5)\cdot (x-5)^2$ and
$\mut_1=1.5$. The curves correspond to (from below)
${\tilde k}=0.49\;(={\tilde k}_0),\; 0.55,\; 0.60$.}
\label{fig5}
\end{figure}
\begin{table}[htb]
\begin{center}
\begin{tabular}{||l|r|c||} \hline \hline
$\mut_1$             & $\tilde k_0~~~$ & $\Delta{\tilde\mu}_{cr}$
                                             \\  \hline \hline
$1.3 = \mut^-$~~     &  1.0~~~    &  ~0.06~      \\ \hline
1.4                  &  0.6~~~    &  ~0.01~      \\ \hline
1.5                  &  0.5~~~    &  ~0.002      \\ \hline
1.6                  &  0.3~~~    &  -0.007      \\ \hline \hline
\end{tabular}
\caption{Wave number and critical Zeeman gap 
$\Delta\mu_{cr} = \mu_1 - \mu_2$ for
some values
of the chemical potential $\mu_1$ at $\lt = 1$. The confining potential
is $V(x)=0.05\cdot\theta(x-5)\cdot(x-5)^2$.
Chemical potentials are measured in units
of the cyclotron energy, whereas the dimensionless wave number is
${\tilde k} = k\cdot l_B$.}
\label{tab1}
\end{center}
\end{table}
We have solved this problem numerically only to lowest order in $Q_2$. 
Harmonic potentials of varying strength have been used, 
and several values 
for $\mu_1$ have been
chosen in the allowed interval (as determined in the one-component case).

Fig.\,\ref{fig3} shows the potential $U(x,k_0)$ for one 
specific value of $\mu_1$. The
minimum of
$U$ is located at the edge as expected. The wave number $k_0$ takes a 
non-vanishing value. Thus, the spin vector is rotating when moving along
the edge.

The corresponding density profiles are shown in Fig.\,\ref{fig4}. The figure
shows that the lower component ($\varphi_2$) of ground state 
density is relatively weakly
bound in
the potential $U(x,k)$ for this particular choice of $V(x)$ and $\mu_1$. 
The localization gets sharper for values of $\mu_1$ corresponding 
to larger $k_0$.
A similar
effect is seen in Fig.\,\ref{fig5} when $k$ is changed relative 
to the ground state 
value $k_0$.

Table\,\ref{tab1} shows numerically estimated values of $k_0$ and the
critical Zeeman
gap for different $\mu_1$ close to the lower critical value $\mu^-$ 
(discussed in Sec$.\,$\ref{bulksec}), and for a fixed confining 
potential $V(x)$. When $\mu_1$ is increased towards $\mu_0$, 
both the wave number and the critical Zeeman gap decrease until $\Delta\mu_{cr}$ goes to zero with a finite value of $k_0$. 
A corresponding study with a steeper confining potential shows 
no solution with positive $k_0$ and $\Delta\mu_{cr}$ close to 
$\mu^-$. This indicates that the interval of $\mu_1$ where 
spin textures occur, increases with softening of the 
confining potential. \\
Similarly, one finds solutions with $\Delta\mu_{cr}>0$ 
near the upper critical value $\mu^+$. These solutions are 
characterized by {\em negative} wave numbers whose absolute 
value decreases along with the critical Zeeman energy as 
$\mu_1$ is lowered.

We can summarize our results as follows: For a sufficiently soft 
confining potential and for Zeeman energies smaller than 
some critical
value, the spins at the edge are tilted in the ground state $(n_z\neq 1)$
and rotate
around the $z$-axis with a wave number $k_0$ as one moves along the edge.
When $\mu_1$
is increased from its lowest allowed value $\mu^-$ at fixed confining
potential, both $k_0$ and the critical Zeeman gap
$\Delta \mu_{cr}$ decrease until $\Delta\mu_{cr}$ goes to zero. 
Near the highest allowed value $\mu^+$ the edge is again 
spin textured, with negative $k_0$ and a critical Zeeman gap 
which approaches zero as $\mu_1$ is decreased.
The ranges in $\mu_1$ where the edge is spin textured, 
decrease as the confining potential gets steeper.

As mentioned in Sec.\,\ref{edgesec}, the egde is expected to become 
unstable with respect to a redistribution of the charge at 
soft confining potentials; for $\mu_1$ away from $\mu_0$ this 
reconstruction is expected to take place at steeper $V(x)$ than 
is the case for $\mu_1=\mu_0$. This means that for fixed $\mu_1$ 
texturing may occur only in some intermediate region where the 
confining potential is neither too steep nor too soft.

Since the ``window'' (Eq.(\ref{incompr})) of allowed $\mu_1$ 
depends on the interaction strength $\lt$, the results 
discussed above may change 
quantitatively for other values of $\lt$; here we have only
examined the case $\lt=1$. For this value of $\lt$ we have 
found values for the critical
Zeeman gap (see table\,\ref{tab1})
which are of the same order of magnitude as the Zeeman gap 
reported in real
experimental
situations; typically, $\Delta \mu/\omega_c\sim 1/60$ in QH 
experiments in GaAs
heterojunctions
\cite{karlhede2}.

\subsection{Edge excitations}

When the edge is spin textured one may expect new massless edge modes to
appear;
as argued in Ref.\,\cite{karlhede1}, translational symmetry along the edge is
spontaneously
broken by the phase of $\phi_2$, and one may expect a corresponding
Goldstone mode to
exist (at least at the mean field level). We have examined the edge modes by
considering
quadratic fluctuations in the fields around the ground state configuration
in the same
way as was done for the spinless case.

Again, we assume the fluctuations in the fields to be separable in the
$x$- and
$y$-dependence,
\be
\delta\rho_{\alpha}(x,y,t) &=& \delta\rho_{\alpha}(x) e^{i(qy-\omega t)} \\
\delta\theta_{\alpha}(x,y,t) &=& -i\eta_{\alpha}(x) e^{i(qy-\omega t)} \\
\delta a_{\mu}(x,y,t) &=& \delta a_{\mu}(x) e^{i(qy-\omega t)}
\ee
and thus replacing $\p_y\rightarrow iq$ and $\p_t\rightarrow -i\omega$, one
finds for
the linearized field equations
\be
\kappa\left(\p_x^2-q^2\right) \delta\sqrt{\rho_{\alpha}}
    &=& \left[ \delta a_0 -\omega\eta_{\alpha}
     + 2\kappa \left({\cal A}_y - k_{\alpha}\right)
         \left( \delta a_y - q\eta_{\alpha} \right)
     + 2\lambda\delta\rho \right] \sqrt{\rho_{\alpha}}  \nonumber \\
    &+& \left[ a_0 + V - \mu_{\alpha} + \kappa ({\cal A}_y - k_{\alpha})^2
     + 2\lambda\rho \right]
                  \delta\sqrt{\rho_{\alpha}}
                                                          \label{2RPA1}\\
\p_x\delta a_y - iq\delta a_x
    &=& -2\theta\delta\rho                          \label{2RPA2} \\
\p_x\delta a_0 - i\omega\delta a_x
    &=& 4\kappa\theta\sum_{\alpha} \left[ \delta\rho_{\alpha}
         \left( {\cal A}_y - k_{\alpha} \right)
     + \rho_{\alpha} \left(\delta a_y - q\eta_{\alpha} \right) \right]
                                                         \label{2RPA3}\\
iq\delta a_0 - i\omega \delta a_y
    &=& -4\kappa\theta \sum_{\alpha}\left[ \rho_{\alpha}\delta a_x
     + i\rho_{\alpha}\p_x\eta_{\alpha} \right]               \label{2RPA4}\\
-i\omega\delta\rho_{\alpha}
    &=& 2\kappa \left[ iq\delta\rho_{\alpha}\left( {\cal A}_y 
     - k_{\alpha} \right)
     +  iq\rho_{\alpha} \left( \delta a_y - q\eta_{\alpha} \right)
        \right.    \nonumber \\
    &+& \left. \p_x\left( \rho_{\alpha} \left( \delta a_x + i\p_x\eta_{\alpha}
                      \right) \right)\right]    \label{2RPA5}
\ee
with $k_1=0$ and $k_2 = k$.
As in Eq.(\ref{qexp}) we expand all the $x$-dependent functions in powers
of $q$ and
assume that the phases $\eta_1$ and $\eta_2$ dominate to $0^{th}$ order. In
analogy
with Eq.(\ref{eta0}), one finds
\be
\sum_{\alpha} \rho_{\alpha}\p_x\eta^{(0)}_{\alpha}
  = \p_x\left( \rho_{\alpha} \p_x\eta^{(0)}_{\alpha} \right) = 0
\ee
which implies that $\eta^{(0)}_1$ and $\eta^{(0)}_2$ are constants.

To ${\cal O}(q)$, the last two equations (\ref{2RPA4}) and (\ref{2RPA5}) give
\be
\delta j^{(1)}_{x,tot} = \p_x\delta j^{(1)}_{x,\alpha} = 0,
\ee
and therefore $\delta j^{(1)}_{x,\alpha} = 0$, as follows from the boundary
conditions
on the current deep in the bulk and outside the edge.

The first three equations, (\ref{2RPA1})-(\ref{2RPA3}), again take the form
of those
for variations in the $t$- and $y$-independent (ground state) fields
$(\delta\rho^{(s)}_{\alpha},
\delta a^{(s)}_{\mu}, \delta\mu_{\alpha},\delta k)$ (see Eqs.
(\ref{S1})-(\ref{S4})), provided
we make the following identifications
\be
\delta a^{(1)}_y - \eta^{(0)}_1 &=& \delta a_y^{(s)}
                                                                 \nonumber \\
\delta a_y^{(1)} - \eta^{(0)}_2 &=& \delta a_y^{(s)} - \delta k \nonumber \\
\delta a_0^{(1)} - \frac{\omega}{q}\;\eta^{(0)}_1 &=& \delta a_0^{(s)} 
                 - \delta\mu_1
                                                             \label{ID1} \\
\delta a_0^{(1)} - \frac{\omega}{q}\;\eta^{(0)}_2
       &=& \delta a_0^{(s)} - \delta\mu_2.                       \nonumber
\ee
One should note that in these expressions $\delta k$ is not restricted by
the condition
(\ref{kGS}), only the ground state conditions (\ref{QGS}) are satisfied.
Thus the
variations in the fields depend on three parameters, which we may choose to be
$Q_1$,$Q_2$ and $k$. However, the identifications (\ref{ID1}) 
implicitly give an
additional constraint,
\be
\omega \delta k= -q(\delta \mu_1-\delta\mu_2) = -q\delta\Delta\mu
\ee
and with this condition included, the fields depend only on two parameters,
$Q_1$ and $Q_2$.

The continuity equations for the currents,
\be
\omega\delta Q_{\alpha}=q\delta J_{\alpha}
\ee
give two linear, homogenous equations in the two variables which determine
the (low
frequency) normal modes. The equations have the form
\be
M_{\alpha \beta} \delta Q_{\beta}=0
\label{M}
\ee
with
\be
M_{\alpha \beta} 
 = (\frac{\omega}{q}+\Delta \mu_k)(\frac{\omega}{q}\delta_{\alpha \beta}
 - J_{\alpha \beta}) + J_{\alpha k}\Delta\mu_{\beta},
\ee
where we have introduced the notations
\be
\Delta \mu_k=\frac{\p \Delta \mu}{\p k} \;\;\;\; 
\Delta \mu_{\alpha}=\frac{\p \Delta\mu}{\p Q_{\alpha}} \nonumber \\
J_{\alpha k}=\frac{\p J_{\alpha}}{\p k}\;\;\;\;\;\;\; 
J_{\alpha \beta}=\frac{\p J_{\alpha}}{\p Q_{\beta}}.
\ee
Eq.(\ref{M}) corresponds to Eq.(\ref{disp1}) for the one-component case,  
which
there gave $\omega/q$ in terms of derivatives of $J$ and $\mu$ with respect
to $Q$. Here
the matrix equation gives a secular equation which is third order in
$\omega/q$,
\be
\left(\frac{\omega}{q}\right)^3+A\left(\frac{\omega}{q}\right)^2  
 + B\frac{\omega}{q}+C=0
\label{seceq}
\ee
with
\be
A &=& \Delta\mu_k-J_{11}-J_{22}                \nonumber \\
B &=&  J_{11}J_{22}-J_{12}J_{21}
   -  \Delta\mu_k(J_{11}+J_{22})
   +  \Delta\mu_1 J_{1k}
   +  \Delta\mu_2 J_{2k}                        \nonumber \\
C &=& \Delta\mu_k(J_{11}J_{22} - J_{12}J_{21})
   -  \Delta\mu_1(J_{1k}J_{22} - J_{2k}J_{12})
   -  \Delta\mu_2(J_{2k}J_{11} - J_{1k}J_{21}).
\ee
The parameters of this equation are determined by the solutions of the
(ground state)
equations (\ref{S4}). We do not have any explicit expressions for the
parameters, but
one should note that they are not all independent, due to relations like
(\ref{mu1J}).

The fact that (\ref{seceq}) is third order in $\omega/q$ indicates that
with spin
effects included there are three massless chiral edge modes (rather than
two). We
cannot give the velocities of these modes in explicit form, but we 
note that the
expressions (\ref{2mu(V)}) and (\ref{mu1J}), which relate the total current
to the
confining potential, imply that the velocities can be written as
\be
\frac{\omega}{q} = \frac{1}{eB} \int_0^{\infty} dx\; V'(x) 
                   \frac{\delta\rho}{\delta  Q}.
                                                    \label{Vdisp2}
\ee
where $\delta \rho$ and $\delta Q$ are the variations in total charge
density and
integrated charge, respectively, for each of the normal 
modes (provided $\delta Q \neq 0$). When the
variation in
the charge density is located at the edge this shows that the velocities of
all the
modes can be written in the general form $-<E>/B$, with $<E>$ as an averaged
value of
$E=-V'/e$ at the edge. However, the $E$-field will be averaged with
functions which in
general are different for each of the three modes.


\section{Discussion}               \label{discsec}

The Chern-Simons Ginzburg-Landau theory gives a mean field description of the
fractional quantum Hall effect which reproduces correctly several of its main
features. Here we have examined, in particular, edge effects in this
framework, both
by analytical and numerical methods. The starting point has been the
observation that if
the chemical potential is varied within a certain interval there is no
change in the
uniform bulk density, but a change in the edge profile is introduced. Thus,
there is a
set of ground states parametrized by the edge charge. The edge current also
varies with
the chemical potential, and we have found that it changes sign for the
value of the
chemical which minimizes the potential energy in bulk. This value can be
considered as
corresponding to the central point of the Hall plateau, the point where the
uniform
charge density of the electrons neutralizes the (constant) background charge.

The low-frequency edge modes have been examined in terms of quadratic
fluctuations in
the fields and with an expansion in the wave number $q$ for fluctuations
propagating
along the edge. As expected (and previously shown for an infinitely steep
confining
potential) there exists in the spinless case one gapless chiral edge mode.
To first
order in $q$ the density profile of this mode is the same as for an
(infinitesimal)
variation in the ground state profile introduced by a variation in the
charge (and the
chemical potential). An expression for the velocity of propagation has been
found in
terms of the confining potential; it has the form of an averaged drift velocity
$E/B$, averaged over the edge with the density profile of the gapless mode.

The edge effects of a fully polarized system with spin have been examined
in a similar
way. In this case there will be a spin texturing of the edge 
if the confining potential is sufficiently soft and if the Zeeman
energy is
not too large. The ground state then depends on two edge charges, one for
each spin
component, and we have examined the case were the second charge
(corresponding to the
spin down component) is small. The critical Zeeman gap has been examined
numerically as
a function of the chemical potential (for a particular value 
of the interaction
strength). It is different from zero in intervals of 
$\mu_1$ near the upper and lower bounds $\mu^+$ and $\mu^-$ 
and decreases when $\mu_1$ approaches $\mu_0$
from either side.
The width of these intervals increases with softening of the confining  
potential.

Also in this case there are gapless edge modes which can be related to
variations in the
ground state fields introduced by variations in the charges. However, the
variations
involve changes in the wave number $k$ of the spin down component which are not
restricted by the ground state condition $\p E/\p k =0$.

The continuity equations for the charges give a third order equation for
the mode
frequencies. This indicates the presence of three chiral modes. However,
the equation
has been given in terms of a set of undetermined parameters. It will be of
interest to
examine further these parameters to see if there are more relations or
constraints than have been established in this paper. Also a numerical
study of these
modes is an interesting subject for further research.

The edge modes have here been studied as quadratic fluctuations about the {\em
mean field} ground state. A question which deserves further study is whether
quantum effects will qualitatively change this picture. The reason that this
cannot be
ruled out is the quasi-one dimensionality of the system and the suggested
connection
to spontanous breaking of translational symmetry of the ground state
\cite{hansson1}. As
is well known, spontaneous symmetry breaking is not 
supported in one-dimensional systems due to (quantum) 
fluctuations in the symmetry breaking variable
\cite{mermin1}.

\bigskip

\begin{center}
{\large \bf Acknowledgements}
\end{center}
We wish to thank Jan Myrheim, Serguei Isakov, Hans Hansson and Dmitri
Khveshchenko for useful discussions and comments.

\eject


\begin{thebibliography}{99}
\bibitem{vonklitzing1} K. von Klitzing, G. Dorda and M. Pepper,
                       {\em Phys. Rev. Lett.} {\bf 45} (1980) 494.
\bibitem{tsui1}        D.C. Tsui, H.L. St\"ormer and A.C. Gossard,
                       {\em Phys. Rev. Lett.} {\bf 48} (1982) 1559.
\bibitem{laughlin1}    R.B. Laughlin, 
                       {\em Phys. Rev. Lett.} {\bf 50} (1983) 1395.
\bibitem{zhang1}       S.C. Zhang, T.H. Hansson and S. Kivelson,
                       {\em Phys. Rev. Lett.} {\bf 62} (1988) 82.
\bibitem{zhang2}       S.C. Zhang, 
                       {\em Int. J. Mod. Phys.} {\bf B 6} (1992) 25.
\bibitem{wen1}         X.G. Wen,  
                       {\em Adv. Phys.} {\bf 44} (1995) 405.
\bibitem{TRANSPORT}    P.L. McEuen, A. Szafer, C.A. Richter, B.W. Alphenaar,
                       J.K. Jain, A.D. Stone, R.G. Wheeler and R.N. Sacks,
                       {\em Phys. Rev. Lett.} {\bf 64} (1990) 2062. \\
                       J.K. Wang and V.J. Goldman,
                       {Phys. Rev. Lett.} {\bf 67} (1991) 749; \\
                       {\em Phys. Rev.} {\bf B 45} (1992) 13 479.
\bibitem{nagaosa1}     N. Nagaosa and M. Kohmoto in
                       {\em ``Correlation effects in low-dimensional 
                       systems''},
                       eds. A. Okiji and N. Kawakami,
                       Springer-Verlag, Berlin (1994), p.168.
\bibitem{orgad1}       D. Orgad and S. Levit, 
                       {\em Phys. Rev} {\bf B 53} (1996) 7964.
\bibitem{morinari1}    T. Morinari and N. Nagaosa, 
                       {\em Solid State Comm.} {\bf 100} (1996) 163.
\bibitem{orgad2}       D. Orgad, 
                       {\em Phys. Rev. Lett.} {\bf 79} (1997) 475.
\bibitem{haldane1}     F.D.M. Haldane, 
                       {\em J. Phys.} {\bf C 14} (1981) 2585.
\bibitem{karlhede1}    A. Karlhede, S.A. Kivelson, K. Lejnell and S.L. Sondhi,
                       {\em Phys. Rev. Lett.} {\bf 77} (1996) 2061.
\bibitem{tafelmayer1}  R. Tafelmayer,
                       {\em Nucl. Phys.} {\bf B 396} (1993) 386.
\bibitem{curnoe1}      S. Curnoe and N. Weiss,
                       {\em Int. J. Mod. Phys.} {\bf A 11} (1996) 329.
\bibitem{ezawa1}       Z.F. Ezawa, M. Hotta and A. Iwazaki,
                       {\em Phys. Rev.} {\bf D 44} (1991) 452.
\bibitem{chamon1}       C. de C. Chamon and X.G. Wen,
                       {\em Phys. Rev.} {\bf B 49} (1994) 8227.
\bibitem{lee1}         D.H. Lee and C.L. Kane, 
                       {\em Phys. Rev. Lett.} {\bf 64} (1990) 1313.
\bibitem{sondhi1}      S.L. Sondhi, A. Karlhede, S.A. Kivelson and E.H. Rezayi,
                       {\em Phys. Rev.} {\bf B 47} (1993) 16419.
\bibitem{karlhede2}    A. Karlhede, S.A. Kivelson and S.L. Sondhi,
                       The quantum Hall effect in {\em Correlated Electron
                       Systems},
                       edited by V.J. Emery, World Scientific (Singapore) 
                       (1993). 
\bibitem{hansson1}     T.H. Hansson, {\em private communication}.
\bibitem{mermin1}      N.D. Mermin and H. Wagner,
                       {\em Phys. Rev. Lett.} {\bf 17} (1966) 1133.
\end{thebibliography}
\end{document}